\documentclass{article}
\usepackage[utf8]{inputenc}
\usepackage{geometry}
\usepackage{auxlib}

\usepackage{makecell}

\usepackage{authblk}

\newcommand{\mst}{\mathsf{MST}}
\newcommand{\opt}{\mathsf{OPT}}

\newcommand{\mmcp}{\mathsf{MMCP}}

\newcommand{\scover}{\mathsf{SC}}

\newcommand{\rank}{\textsf{rank}}

\newcommand{\kru}{\mathsf{Kruskal}}

\def\MMCP/{MMCP}

\newcommand{\vpreceq}{\boldsymbol{\preceq\ }}
\newcommand{\vsigma}{\boldsymbol{\sigma}}

\title{Multiagent MST Cover: Pleasing All Optimally via A Simple Voting Rule\footnote{All authors (ordered alphabetically) have equal contributions and are corresponding authors.}}

\author[1]{Bo Li}
\author[2]{Xiaowei Wu}
\author[3,4\thanks{This work was done when the author was a student at Zhejiang University.}]{Chenyang Xu}
\author[5\thanks{This work was done when the author visited University of Macau.} ]{Ruilong Zhang}

\affil[1]{\footnotesize Department of Computing, The Hong Kong Polytechnic University}
\affil[2]{\footnotesize IOTSC, University of Macau}
\affil[3]{\footnotesize Software Engineering Institute, East China Normal University}
\affil[4]{\footnotesize College of Computer Science, Zhejiang University}
\affil[5]{\footnotesize Department of Computer Science, City University of Hong Kong}
\affil[ ]{\texttt{comp-bo.li@polyu.edu.hk, xiaoweiwu@um.edu.mo, xcy1995@zju.edu.cn, ruilzhang4-c@my.cityu.edu.hk}}

\date{}

\begin{document}

\maketitle

\sloppy

\begin{abstract}

Given a connected graph on whose edges we can build roads to connect the nodes, a number of agents hold possibly different perspectives on which edges should be selected by assigning different edge weights. 
Our task is to build a minimum number of roads so that every agent has a spanning tree in the built subgraph whose weight is the same as a minimum spanning tree in the original graph.
We first show that this problem is $\NP$-hard and does not admit better than $((1-o(1))\ln k)$-approximation polynomial-time algorithms unless $\comP = \NP$, where $k$ is the number of agents.
We then give a simple voting algorithm with an optimal approximation ratio.
Moreover, our algorithm only needs to access the agents' rankings on the edges.
Finally, we extend our results to submodular objective functions and Matroid rank constraints.


\end{abstract}

\newpage

\section{Introduction}

Minimum spanning tree (MST) is one of the most fundamental problems in graph theory, whose objective is to find a subgraph in a given graph to span all the nodes with minimum total weight.
It finds wide applications in computer science,
among which a typical scenario is network construction, such as fiber optic network design~\cite{DBLP:conf/or/BachhieslPPWS02}, highway system design~\cite{DBLP:books/daglib/0069809}, FIR filter implementation~\cite{ohlsson2004implementation}, and so on. 
The problem becomes trickier when the system interacts with multiple heterogeneous entities who hold possibly different perspectives on measuring the weights of the edges in the graph.
For example, when a government is planning the traffic in a city, many bureaus, such as water resources, communications, construction, finance, and information industry, are involved, but they may have different concerns and thus hold different opinions on which roads should be built.
Since there may not exist one MST that satisfies all bureaus, we focus on the minimum cost maximum social welfare problem by building the minimum number of roads so that all bureaus are satisfied.
Formally, we are interested in the following optimization problem: Given a graph where different agents may have different weight functions on the edges, how can we select a minimum number of edges so that the induced subgraph contains at least one MST of the original graph for every agent.




Our problem is closely related to the multiple objective minimum spanning tree (MOST) problem, whose goal is to select a spanning tree $T$ in the graph to minimize $(g_1(T),\ldots,g_k(T))$, where each $g_i(\cdot)$ represents a different weight function.
Various multi-dimensional optimality concepts have been investigated in the literature \cite{DBLP:journals/siamcomp/HassinL04,DBLP:journals/orl/HongCP04,DBLP:journals/symmetry/MajumderBBBMKZ22,DBLP:conf/dfg/RuzikaH09}, among which two of the most popular ones are min-sum \cite{DBLP:journals/coap/CorreiaPF21,DBLP:journals/entcs/FernandesMGG19} and min-max \cite{DBLP:journals/mss/DarmannKP09,DBLP:journals/aamas/EscoffierGM13}, i.e., finding a spanning tree $T$ to minimize $\sum_{i\in[k]}g_i(T)$ or $\max_{i\in[k]}g_i(T)$.
Although computing a minimum spanning tree in the single-objective setting is polynomial-time solvable, it appears to be $\NP$-hard for both multi-dimensional optimality concepts. 
Similar problems are also studied in computation social choice, where the objective is to find one spanning tree such that the satisfaction of the least satisfied agent gets maximized under different ``satisfaction'' notions \cite{DBLP:journals/mmor/Darmann16,DBLP:journals/mss/DarmannKP09,DBLP:journals/aamas/EscoffierGM13,DBLP:journals/dam/GourvesMT15}. 
The key difference between the existing research and our work is that we are not targeting a spanning tree; instead, we are allowed to select a spanning {\em subgraph} and aim to please all agents with a minimum number of edges.

\begin{figure}[htb]
    \centering
    \includegraphics[width=13cm]{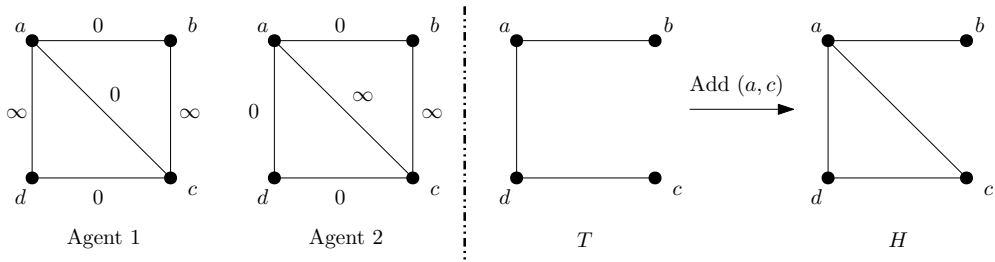}
    \caption{An illustration of the satisfiability issue when the solution is restricted to a spanning tree. 
    }
    \label{fig:motivation}
\end{figure}

Regardless of which multi-dimensional optimality concept is adopted, selecting one spanning tree for all agents can lead to extremely bad situations for some of them. 
We show an example in \cref{fig:motivation}, where $\infty$ denotes a large constant.
There are two agents.
The unique MST of agent $1$ and agent $2$ is $\set{(a,b),(a,c),(c,d)}$ and $\set{(a,b),(a,d),(c,d)}$, respectively.
If we restrict any feasible solution to be a spanning tree, no matter which spanning tree we select, there will always be half of the agents that dislike it very much. 
For example, the selected spanning tree is $T=\set{(a,b),(a,d),(c,d)}$ in the figure.
Then agent $1$ is
depressed as her valuation of the selected spanning tree is $\infty$, much larger than her original MST's value $0$. 
But if we are allowed to add one more edge to the solution, all agents will be satisfied because each of them can find her original MST from the given subgraph $H=\set{(a,b),(a,c),(a,d),(c,d)}$.


In the urban traffic planning example, and more practical applications, building a spanning tree is more economic than building a spanning subgraph with more edges. 
However, it may make some agents unsatisfied, and thus, lead to a poor social welfare.
On the top of maximizing the social welfare, we want to find an economic way to please all agents, which motivates our model.

\subsection{Our Contributions}
We study the multiagent MST cover problem (MMCP), where the goal is to select a minimum set of edges from a given graph that covers a minimum spanning tree for every agent.
In a nutshell, our main results can be summarized as follows.

\medskip

{\bf Main Results.}
We show that the multiagent MST cover problem is as hard as set cover and design an optimal logarithmic approximation algorithm, which also works when the algorithm only accesses the ordinal information of agent preferences. 
Further, the proposed algorithm can be extended to submodular objective functions and Matroid rank constraints.

\medskip

In a bit more details, we first show that the optimization problem is $\NP$-hard via a reduction from the set cover problem, and does not admit better than $((1-o(1))\ln k)$-approximate polynomial-time algorithms unless $\comP = \NP$, where $k$ is the number of agents.
Interestingly, a special case of the decision problem---deciding whether there is a spanning tree that is minimum for all agents, can be solved in polynomial time. 

We then design a polynomial-time algorithm that achieves the optimal approximation ratio. 
Intuitively, in each round of our algorithm, we ask the agents which edge should be added to the solution and each agent votes for the ones that make the solution  ``closer'' to one of her minimum spanning trees. 
The algorithm adds the edge that receives the most votes and stops when all agents find a minimum spanning tree in the solution. 
Technically, we define a progress function to evaluate how close our partial solution is to satisfying all agents, and then show that our algorithm is equivalent to adding
edges to maximize the marginal progress.
By proving that the progress function is submodular, we obtain the logarithmic approximation.

Our algorithm is robust in two senses.
First, it does not need to access the cardinal edge-weights of the agents; as long as we have the ordinal rankings from the agents, the approximation ratio is guaranteed. 
This makes our algorithm practically useful as it may be hard for the government to gather all the accurate numerical values from the agents.
Second, our algorithm can be extended to more complicated scenarios.
For example, due to the heterogeneity of the edges (e.g., lengths and geologic structures)
the construction cost of different edges can be different for the government. Thus instead of selecting the minimum number of edges, the government wants to minimize the total construction cost of the selected edges.
We show that our algorithmic idea still works in the case that the construction cost function (over edges) is monotone and submodular. 
In addition to the generalized objective, we also extend the minimum spanning tree constraints to maximum Matroids rank, where our results also apply.

\subsection{Other Related Works}

\paragraph{Multi-objective Spanning Tree.}
Finding a minimum spanning tree is a fundamental problem in multi-objective optimization~\cite{DBLP:books/daglib/0022057}.
See~\cite{DBLP:conf/dfg/RuzikaH09} for a survey.
Typically, we are given a connected undirected graph $G(V,E)$.
Each edge $e\in E$ has a positive weight vector $\vw(e)=(w_1(e),\ldots,w_k(e))$.
For every spanning tree $T$, we have a weight vector $\vw(T)=(w_i(T))_{i\in[k]}$ with $w_i(T)=\sum_{e\in E}w_i(e)$.
The goal is to compute a spanning tree $T$ such that $T$ is Pareto optimal.
This problem has been shown to be $\NP$-hard even if the weight vector only has two coordinates~\cite{DBLP:journals/anor/HamacherR94} in which the problem is called the bi-objective spanning tree minimization problem. 
The literature mainly looks into approximation algorithms and exact exponential time algorithms.
Bi-objective spanning tree minimization has also been studied extensively~\cite{DBLP:journals/itor/AmorosiP22,DBLP:journals/rairo/SousaSA15}.
For approximation algorithms, a fully polynomial-time approximation scheme (FPTAS) is proposed by \cite{DBLP:conf/focs/PapadimitriouY00}.
For exact algorithms, most existing algorithms are widely used evolutionary methods in multi-objective optimization~\cite{DBLP:journals/ijcse/LaiX19,DBLP:journals/eaai/PrakashPS20}.

\paragraph{Fair Allocation with Public Goods.}
Our problem shares some similarities with the problem of fair allocation of public goods.
Our problem and technique are different since we consider specific forms of aggregating
the objectives.
Fair allocation with private goods is an explosive area in which goods are required to be assigned to some agent, and other agents do not obtain utilities.
This differs from public goods, where all agents obtain utilities when some goods are selected.
See~\cite[sec.~4]{conitzer2019algorithmically} for a brief discussion between public and private goods.
Fair allocation of public goods is also closely related to voting in combinatorial domains (see book~\cite{DBLP:reference/choice/LangX16} for an overview).
\textcite{DBLP:conf/sigecom/ConitzerF017} first use ``public" to distinguish the previous ``private" goods.
They aim to select a set of goods to satisfy the fairness notion of propositional share. In \cite{DBLP:conf/sigecom/FainM018}, they propose a new fairness notion for public goods which generalizes the propositional share.
They also consider some constraints of goods, i.e., the select goods must satisfy some properties, like matching, packing, and matroid.
Later, \textcite{DBLP:conf/fsttcs/GargKM21} investigated the relationship between public goods and private goods when the objective is to maximize Nash social welfare.
\textcite{DBLP:conf/aaai/FluschnikSTW19} also aim to maximize Nash social welfare under the multiagent knapsack problem.

\paragraph{Partial Information Setting.}
In the current work, due to the practical concern when it is hard for the decision maker to elicit agents' complete information of their preferences, we only assume the algorithm has access to their ordinal rankings.
Similar settings have been widely investigated in computational social choice. 
\citet{conf/cia/ProcacciaR06} first studied elections under the Borda voting rule, and introduced {\em distortion} which is used to bound the worst-case deterioration of the optimal objective with limited access of preferences.
Later, \citet{journals/ai/CaragiannisP11} and \citet{journals/jair/CaragiannisNPS17} continued to study the same problem under the Plurality rule.
There is also a line of works, e.g., \citet{journals/ai/AnshelevichBEPS18, conf/ec/MunagalaW19,conf/focs/Gkatzelis0020}, that studies the  metric social choice with partial information when the alternatives are points in a metric space. 
We refer the readers to \cite{conf/ijcai/AnshelevichF0V21} for a detailed survey.

\subsection{Roadmap}

The formal definition of multiagent MST cover is stated in \cref{sec:model}.
\cref{sec:hardness} proves that the problem is set-cover hard, while \cref{sec:prefect-cover} gives a simple and efficient algorithm to determine whether there is a spanning tree that is an MST for all agents.
In~\cref{sec:approx}, we present our optimal logarithmic approximation algorithm, and then we show that our approximation algorithm can be extended into the more general case in \cref{sec:extension}.
Finally, we conclude this paper and discuss some interesting future works in \cref{sec:conclusion}.

\section{Preliminaries}
\label{sec:model}

Consider a connected undirected graph $G(V,E)$ with $n$ nodes and $m$ edges.
There is a set $K$ of $k$ agents, where each agent $i\in K$ has a preference $\preceq_i$ over the edges $E$.
Denote by $\vpreceq=(\preceq_1,\ldots, \preceq_k)$ the preference profile of all agents.
For any two edges $a,b\in E$, $a \prec_i b$\footnote{Since agents prefer smaller cost in our model, we use $a \prec_i b$ to denote that $w_i(a) < w_i(b)$, which might be different from other literature of computational social choice, in which $a \prec_i b$ means that agent $i$ prefers $b$ than $a$.} means that from the view of agent $i$, edge $a$ is cheaper than edge $b$, while $a \sim_i b$ represents the case that agent $i$ is indifferent between $a$ and $b$.
Clearly, $\preceq_i$
can be denoted by $\kappa_i$ equivalence classes
$\cE(\preceq_i):=(E_i^1,\ldots,E_i^{\kappa_i})$
(in increasing order of costs), where each $E_i^j$ contains the edges among which agent $i$ is indifferent, and $\kappa_i$ is the number of her equivalence classes.  
Accordingly, for any two edges $a\in E_i^j$ and $b\in E_i^{\ell}$, $a\sim_i b$ implies $j = \ell$ and $a\prec_i b$ implies $j < \ell$.

Agents may have arbitrary cardinal weight functions on the edges as long as they are {\em consistent} with $\vpreceq$.
Specifically, agent $i\in K$ has weight function $w_i: E \to \R_{\geq 0}$ that is consistent with $\preceq_i$ if $w_i$ satisfies: $w_i(a) < w_i(b)$ (resp. $w_i(a)=w_i(b)$) if and only if $a \prec_i b$ (resp. $a\sim_i b$) for any $a,b\in E$.
We denote by $\vw=(w_1,\ldots,w_k)$ the weight profile of all agents.
In our model, the algorithms only have access to the preferences $\vpreceq$, but not the weight functions $\vw$. 

Given any graph $G(V,E)$ and a set of edges $S\subseteq E$, let $G[S] := (V,S)$ be the subgraph of $G$ induced by $S$. 
Under the weight function $w$, a {\em minimum spanning tree} (MST) of $G$ is an induced subgraph $G[S]$ that is connected, and $w(S):=\sum_{e\in S}w(e)$ is minimized.
When the context is clear, we also use $S$ to denote the induced subgraph $G[S]$.
Let $\mst(G, w)$  and $\mst(G,\preceq)$ be the set of all MSTs of $G$ under weight function $w$  and preference profile $\preceq$, respectively, where a spanning tree $T$ is minimum under $\preceq$ if $T$ is an MST under any weight functions that are consistent with $\preceq$.
The following lemma captures the equivalence between $\mst(G,\preceq)$ and $\mst(G,w)$.

\begin{lemma}
For any preference profile $\preceq$ and weight function $w$ consistent with it, $\mst(G,w)=\mst(G,\preceq)$.
\label{lem:mst_prec_weight}
\end{lemma}


\begin{proof}
Consider an arbitrary preference $\preceq$, let $\cW$ be the set of weight functions that are consistent with the preference $\preceq$.
To show \cref{lem:mst_prec_weight}, it is sufficient to show that $\mst(G,w_i)=\mst(G,w_j)$ for any $w_i,w_j\in\cW$.
Let $T$ be an arbitrary MST in $\mst(G,w_i)$.
We show that $T$ is also in $\mst(G,w_j)$.
Assume the contrary that $T\notin\mst(G,w_j)$.
This implies that there is an edge $e\in E\setminus T$ such that (\rom{1}) $T\cup\set{e}$ contains a cycle $C$; (\rom{2}) there is another edge $e'$ in the cycle such that $w_j(e)<w_j(e')$.
Since $w_j$ is consistent with $\preceq$ and $w_j({e})<w_j({e'})$, we know that $e\prec e'$.
Thus, we have $w_i({e})<w_i({e'})$ since $w_i$ is also consistent with $\preceq$.
Hence, we have $w_i(T\setminus\set{e'}\cup\set{e})<w_i(T)$.
Therefore, $T$ is not an MST under the weight function $w_i$, which contradicts our assumption.
By the symmetry of $i$ and $j$, we have $\mst(G,w_i)=\mst(G,w_j)$.
\end{proof}

For simplicity, denote $\mst(G,\preceq_i)$ by $\mst_i$ for each agent $i\in K$.
For a positive integer $x$, we use $[x]$ to denote $\set{ 1,2,\ldots,x }$.
Based on \cref{lem:mst_prec_weight}, checking whether a spanning tree is an MST under some preference $\preceq$ can be done in polynomial time.

A feasible solution of the {\em multiagent MST cover problem} (\MMCP/) is a set of edges $H$ such that for any agent $i$, there exists $T\subseteq H$ which is an MST for agent $i$.
Namely, any agent has at least one MST that is covered by $H$. 
Call such an edge set an \emph{MST cover} of all agents.
Our goal is to find an MST cover for all agents with the minimum size.
Note that in \cref{sec:extension}, we also consider the weighted version of \MMCP/, in which we have a cost function $c:2^E \to \R_{\geq 0}$ on the edges, and the objective is to find an MST cover $H$ of all agents with minimum cost $c(H)$.

By \cref{lem:mst_prec_weight}, given any connected undirected graph $G(V,E)$ with a preference $\preceq$ over $E$, it is well-known that Kruskal's algorithm~\cite{kruskal1956shortest} computes an MST of $G$.
For completeness, we restate the algorithm in \cref{alg:kruskal}.
Let $\kru(G,\preceq)$ be the returned MST by Kruskal's algorithm given graph $G$ and preference $\preceq$.

\begin{algorithm}[htb] 
\caption{Kruskal's Algorithm}
\label{alg:kruskal}
\begin{algorithmic}[1]
\REQUIRE A connected undirected graph $G(V,E)$; a preference $\preceq$ over $E$.
\ENSURE An MST $T$ of $G$ under $\preceq$.
\STATE $T\leftarrow\emptyset$; $i\leftarrow 1$.
\STATE Let $(e_1,e_2,\ldots,e_m)$ be the list of edges sorted by $\preceq$ (break tie arbitrarily).
\WHILE{$i\leq m$}
\IF{there is no cycle in $T\cup\set{e_i}$}
\STATE $T\leftarrow T\cup\set{e_i}$.
\STATE $i\leftarrow i+1$.
\ENDIF
\ENDWHILE
\RETURN $T$.
\end{algorithmic}
\end{algorithm}

\section{Hardness Results}
\label{sec:hardness}

In this section, we show that \MMCP/ is $\NP$-hard to approximate within a factor better than $\ln k$ even when (\rom{1}) all weight functions $w_1(\cdot),\ldots,w_k(\cdot)$ are given; (\rom{2}) all weight functions are binary, i.e., for any $i\in K$ and  $e\in E$, $w_i(e)\in \set{0,1}$.
We refer to this special case as {\em \MMCP/ with binary weights}.
To warm up, we first introduce a reduction to show the $\NP$-hardness. 

\begin{theorem}\label{thm:hardness}
\MMCP/ with binary weights is $\NP$-hard.
\end{theorem}
\begin{proof}
The NP hardness is proved by a reduction from Set Cover (SC) to the \MMCP/ problem in which $w_i(e) \in \{0,1\}$ for any $i\in K$ and $e\in E$. 
In the set cover problem, we are given a ground element set $U = \{ e_1,\ldots,e_p \}$ and a subset collection $S = \{ s_1,\ldots,s_q \}\subseteq 2^U$. The goal is to find a minimum number of subsets in $S$ whose union covers $U$.

Considering an arbitrary SC instance with $p$ elements and $q$ subsets, we construct an \MMCP/ instance with a graph $G(V,E)$ and $p+1$ agents as follows.
The graph $G(V,E)$ contains $q+1$ nodes and $2q-1$ edges.
Let $V = \{ a, b_1,\ldots,b_q \}$.
For each $i\in[q]$, we connect $a$ and $b_i$, and refer to edge $(a,b_i)$ as an \emph{up-edge}.
For each $i\in [q-1]$, connect $b_i$ and $b_{i+1}$, and call this edge a \emph{down-edge}.
Now we define the weight functions of the $(p+1)$ agents. 
First, let all the down-edges have weight $0$ for all agents.
Then, for each agent $j\in [p]$, let $w_j(a,b_i) = 0$ if element $e_j\in s_i$; $w_j(a,b_i) = 1$ otherwise.
Note that by this construction, the value of MST for every agent $j\in [p]$ is $0$.
Let the last agent have weight $1$ on all up-edges, and thus the MST's value for agent $p+1$ is $1$.
It is easy to see that the reduction can be implemented in polynomial time.
See \cref{fig:hardness} for an illustration.

\begin{figure}[htb]
    \centering
    \includegraphics[width=13.5cm]{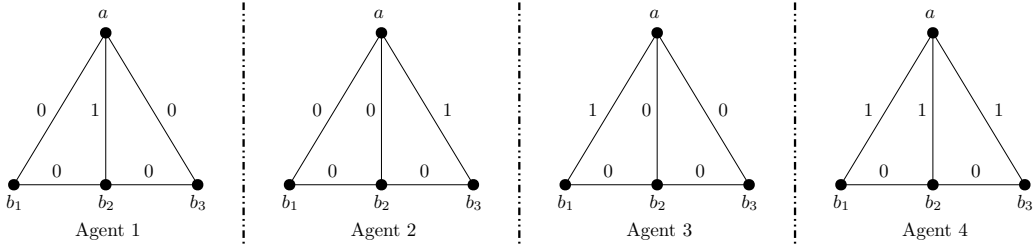}
    \caption{Illustration for the reduction. The set cover instance $(S,U)$ is as follows: $U=\set{e_1,e_2,e_3}$ and $S=\set{s_1,s_2,s_3}$ with $s_1=\set{e_1,e_2}$, $s_2=\set{e_2,e_3}$ and $s_3=\set{e_1,e_3}$. 
    We construct a graph $G(V,E)$ with $V=\set{a,b_1,b_2,b_3}$ and $E=\set{(a,b_1),(a,b_2),(a,b_3)}\cup\set{(b_1,b_2),(b_2,b_3)}$. There are four agents with distinct weight vectors shown in the figure.}
    \label{fig:hardness}
\end{figure}
Observe that any MST cover $H$ of agents must contain all down-edges; otherwise, $H$ does not cover any MST of the last agent. 
Moreover, $H$ must contain at least one up-edge $e$ with $w_j(e)=0$ for any $j\in [p]$ due to a similar argument. 
By interpreting picking edge $(a,b_i)$ in \MMCP/ as selecting set $s_i$, there is a one-to-one correspondence between feasible solutions to these two problems.
Thus, computing the subgraph cover $H^*$ corresponds to finding an optimal solution for the set cover problem, and the hardness result follows.
\end{proof}

Now we build on the reduction above to show an inapproximability result.

\begin{theorem}
Unless $\comP = \NP$, any polynomial-time algorithm for \MMCP/ with binary weights has approximation ratio at least $(1-o(1))\ln k$, where $k$ is the number of agents.
\end{theorem}
\begin{proof}
We alter the reduction from SC to \MMCP/ in the proof of \Cref{thm:hardness} to show the hardness of approximation.
For a set cover instance $(U,S)$ with $p$ elements and $q$ subsets, the graph is constructed in a similar way as before, but with sufficiently many copies of node $a$.

Formally speaking, let $V=\set{a_1,\ldots,a_{h\cdot(q-1)},b_1,\ldots,b_q}$ and
\begin{equation*}
    E = \set{ (a_t,b_i) : t\in[h\cdot(q-1)], i\in[q] } \cup \set{ (b_i,b_{i+1}) : i\in[q-1] },
\end{equation*}
where $h$ is an integer that will be defined later. 
Namely, we create $h\cdot(q-1)$ copies of node $a$ and connect each of these copies to every node $b_i$ ($i\in [q]$). 
There are $p+1$ agents, and the weight functions are the same with those in \cref{thm:hardness}. 
That is the weights of all down-edges are 0 for all agents; 
for any agent $j\in [p]$ and $t\in [h(q-1)]$, $w_j(a_t,b_i)=0$ if $e_j\in s_i$, and $w_j(a_t,b_i)=1$ otherwise;
for the last agent $p+1$, all up-edges have weight $1$. 
See \cref{fig:lowerbound} for an illustration.

\begin{figure}[htb]
    \centering
    \includegraphics[width=10cm]{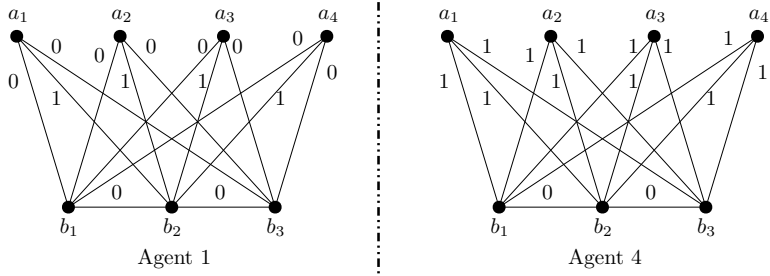}
    \caption{An illustration for the reduction for the hardness of approximation. The set cover instance $(U,S)$ has $U=\set{e_1,e_2,e_3}$ and $S=\set{s_1,s_2,s_3}$ with $s_1=\set{e_1,e_2}$, $s_2=\set{e_2,e_3}$ and $s_3=\set{e_1,e_3}$. 
    We construct a graph $G(V,E)$ with $V=\set{a_1,\ldots,a_4,b_1,b_2,b_3}$ and $E=\bigcup_{t=1,2,3,4}\set{(a_t,b_1),(a_t,b_2),(a_t,b_3)}\cup\set{(b_1,b_2),(b_2,b_3)}$. There are four agents with distinct weight vectors. In the figure, we only show the weight of agent $1$ and $4$.}
    \label{fig:lowerbound}
\end{figure}

Similar to \cref{thm:hardness}, the constructed \MMCP/ instance has the following two properties.
\begin{itemize}
    \item For any feasible solution to the \MMCP/ instance, the neighbours of every $a_t$, where $t\in [h(q-1)]$, form a feasible solution to the original SC instance.
    
    \item For any optimal solution of the \MMCP/ instance, the neighbours of every $a_t$, where $t\in [h(q-1)]$, form an optimal solution of the original SC instance.
\end{itemize}

Suppose that there exists an algorithm $\cA$ with approximation ratio $(1-\epsilon)\ln |K|$, for some constant $\epsilon>0$.
We show that this contradicts to the inapproximability bound of set cover.
Recall that in the constructed \MMCP/ instance, $|K| = p+1$ and $|V|=q+1$.
Let $\cA(\mmcp)$ and $\opt(\mmcp)$ be the solution returned by algorithm $\cA$ and the optimal solution, respectively. 
We have
\begin{equation*}
    |\cA(\mmcp)| \leq (1-\epsilon) \ln (p+1) \cdot |\opt(\mmcp)|.
\end{equation*}

According to the first property, the neighbourhood of any node $a_t$ in $\cA(\mmcp)$ is a feasible solution of the original SC instance. Pick the node $a$ with the minimum degree and let $\cA(\scover)$ be its set of neighbours. 
Clearly, 
\begin{equation*}
    h(q-1)\cdot |\cA(\scover)| + q - 1 \leq |\cA(\mmcp)|.
\end{equation*}

Use $\opt(\scover)$ to denote an optimal solution to the set cover instance. Then we connect $\opt(\mmcp)$ to $\opt(\scover)$:
\begin{equation*}
    |\opt(\mmcp)| = h(q-1) \cdot |\opt(\scover)| + q-1.
\end{equation*}

Combining the above three equations, we have
\begin{equation*}
    h(q-1)\cdot |\cA(\scover)| + q-1 \leq (1-\epsilon) \ln (p+1) \cdot (h(q-1)\cdot |\opt(\scover)| + q-1)
\end{equation*}

Finally, reordering the inequality gives
\begin{equation*}
    |\cA(\scover)| \leq (1-\epsilon) \ln (p+1) \cdot |\opt(\scover)| + \frac{(1-\epsilon)\ln(p+1)-1}{h},
\end{equation*}
which is a contradiction to the $(1-o(1))\ln p$ lower bound of set cover~\cite{DBLP:journals/jacm/Feige98} under the assumption that $\comP\neq \NP$, when $h$ is sufficiently large, e.g., $h\geq 1/(2\epsilon)$.
\end{proof}

\section{Perfect MST Cover}
\label{sec:prefect-cover}

Although \MMCP/ is $\NP$-hard and has an inapproximability of $(1-o(1))\cdot \ln k$, we find that deciding whether there is a spanning tree that is an MST for all agents can be decided efficiently.
We refer to this common MST of all agents as a {\em Perfect MST Cover}. 
This section shows a polynomial time algorithm to decide the existence of a perfect MST cover. 

Given any agent $i$ with her preference $\preceq_i$, recall that $\cE(\preceq_i)=(E_i^1,\ldots,E_i^{\kappa_i}$) is an ordered partition of $E$ generated by $\preceq_i$.
Let $\sigma_i:E\to [m]$ be a rank function for agent $i$, where $\sigma_i(e)=j$ if $e\in E_i^j$, i.e., $\sigma_i(e)$ is the rank of edge $e$ for agent $i$.
Fix an arbitrary ordering of agents and let $\vsigma(e) = ( \sigma_1(e),\ldots,\sigma_k(e) ) \in [m]^{k}$ be the rank vector of edge $e$.
Note that $\vsigma$ induces a lexicographical ordering on the edges, i.e., $\vsigma(e_1) < \vsigma(e_2)$ if and only if there exists $i\in K$ such that $\sigma_j(e_1) = \sigma_j(e_2)$ for all $j<i$ and $\sigma_i(e_1) < \sigma_i(e_2)$. 

\begin{algorithm}[htb] 
\caption{Computing a Perfect MST Cover}
\label{alg:common-MST}
\begin{algorithmic}[1]
\REQUIRE A connected undirected graph $G(V,E)$; Agent set $K$ and preferences $\vpreceq =(\preceq_1,\ldots,\preceq_k)$.
\ENSURE A spanning tree $T$ of $G$.
\STATE Sort all edges in $E$ by the rank vector $\sigma$ in lexicographical order.
\STATE Let $\preceq$ be the ordering of the edges after sorting.
\STATE $T\leftarrow \kru(G,\preceq)$.
\RETURN $T$.
\end{algorithmic}
\end{algorithm}

The description of the algorithm is given in \cref{alg:common-MST}. Clearly, \cref{alg:common-MST} is a polynomial time algorithm because sorting all edges by the rank vector $\sigma$ takes $\cO(km\log m)$ time by aggregating the preferences and Kruskal's Algorithm runs in $O(m \log n)$.
The following theorem shows that if an instance admits a perfect MST cover, then \cref{alg:common-MST} computes one.

\begin{theorem}
Given any instance of \MMCP/, \cref{alg:common-MST} returns a perfect MST cover $T$ if and only if the instance admits one. 
\label{thm:perfec_cover}
\end{theorem}
\begin{proof}


Suppose the instance admits a perfect MST cover.
Consider any weight function $\vw=(w_1,\ldots,w_k)$ that is consistent with the given preference $\vpreceq= (\preceq_1,\ldots,\preceq_k)$. 
We show that $T$ is a perfect MST cover under the weight function $\vw$. 
Let $w_i(T)$ be the total weight of $T$ for agent $i\in K$. 
According to the statement of \cref{alg:common-MST}, $T$ is an MST under the preference $\preceq'$. Thus, among all spanning trees of the graph, $T$ has the minimum weight vector $(w_1(T),\ldots,w_k(T))$ (lexicographical order). 
So if there exists a perfect MST cover $T'$, we have 
\begin{equation*}
    (w_1(T'),\ldots,w_k(T')) \geq  (w_1(T),\ldots,w_k(T)).
\end{equation*}

On the other hand, due to the definition of a perfect MST cover, for any $i\in K$, we have
\begin{equation*}
    w_i(T)\geq w_i(T').
\end{equation*}

Thus, for all $i\in K$, we have $w_i(T)=w_i(T')$, implying that $T$ is a perfect MST cover.
\end{proof}

\begin{remark}
It deserves to note that there are other ways to compute a perfect MST cover besides implementing \cref{alg:common-MST}.
For example, we can arbitrarily assign each agent a weight function $w_i(\cdot)$ that is consistent with her preference, and then sum up all agents' weights $\bar{w}(\cdot) = \sum_i w_i(\cdot)$.
We can compute an MST on $\bar{w}(\cdot)$ (denoted by $T$), and prove that there is a common MST if and only if 
$T\in \mst_i$ for any agent $i$.
\end{remark}

\section{Multi-Round Plural Voting Algorithm}
\label{sec:approx}

This section gives an $\cO(\ln k)$-approximation algorithm (\cref{alg:approx}) and shows the following main result (\cref{thm:submodular_algo}).
Note that the approximation ratio is asymptotically optimal, by the hardness result  we have shown in \cref{sec:hardness}.

\begin{theorem}\label{thm:submodular_algo}
For every instance of \MMCP/, \cref{alg:approx}
computes an $\cO(\ln k)$-approximation solution in polynomial time, where $k$ is the number of agents.
\end{theorem}


\paragraph{Algorithmic Framework.}
Our algorithm runs in rounds and adds exactly one edge to the solution $H$ (which is initially empty) in each round until the solution is feasible, i.e., it contains an MST for every agent.
At the beginning of each round, every agent specifies a candidate edge set $S_i\subseteq E\setminus H$ from the unselected edges.
Intuitively, the candidate edges in $S_i$ are those agent $i$ prefers to select, given the current partial solution $H$.
Then, our algorithm chooses the edge that appears in most candidate edge sets (break ties arbitrarily), e.g., the edge that receives the most ``votes'', and adds it to the solution set $H$.
In other words, the algorithm repeats the following steps until $H$ becomes feasible:
\begin{itemize}
    \item each agent $i\in K$ specifies a candidate edge set $S_i\subseteq E\setminus H$;
    \item let $e^*\in \arg\max_{e\in E\setminus H} |\{ i\in K : e\in S_i \}|$ be an edge that receives the most votes;
    \item add $e^*$ to the solution: $H \gets H\cup \{e^*\}$.
\end{itemize}

Apparently, the crucial part of the algorithm is how to define the candidate edge set for each agent in each round.
In our algorithm, we introduce a \emph{progress function} $f_i : 2^E \to \{ 0,1,\ldots,n-1 \}$ for each agent $i\in K$ that measures how much progress the current solution $H$ has made for her, e.g., the larger $f_i(H)$ is, the more progress has been made for agent $i$.
The candidate edges for agent $i$, are naturally those whose inclusion to $H$ makes $f_i(H)$ larger.

%

\begin{definition}[Progress Function]
Given any agent $i\in K$ and any edge set $H\subseteq E$, the progress function $f_i$ of agent $i$ is defined as $f_i(H):=\max_{T\in\mst_i} \{ |H \cap T| \}$.
\label{def:progress_function}
\end{definition}

In other words, $f_i(H)$ is the maximum number of edges from $H$ that can simultaneously appear in some MST of agent $i$. 
Note that the function $f_i$ is monotone and $f_i(H) = n-1$ if and only if the set $H$ contains an MST for agent $i$. 
Therefore, $(n-1) - f_i(H)$ indicates the minimum number of edges that need to be added into $H$ so that the solution becomes feasible to agent $i$.
The closer $f_i(H)$ is to $n-1$, the agent is more satisfied with the partial solution $H$.

\paragraph{Example.} 
Let us consider the example from \cref{fig:motivation}.
Note that the unique minimum spanning tree for agent $1$ and $2$ are $T_1=\set{(a,b),(a,c),(c,d)}$ and $T_2=\set{(a,b),(a,d),(c,d)}$, respectively.
If the partial solution $H=\set{(a,b),(a,c)}$, then $f_1(H)=2$ (resp. $f_2(H)=1$) since $\abs{T_1\cap H}=2$ (resp. $\abs{T_2\cap H}=1$).
If we add one more edge $(a,d)$ to $H$, the new partial solution becomes $H'=\set{(a,b),(a,c),(a,d)}$.
In this case, we have $f_1(H')=2$ and $f_2(H')=2$.
Since $f_2(H\cup\set{(a,d)})>f_2(H)$, edge $(a,d)$ is among the candidate edges of agent $2$, given the partial solution $H$.

\subsection{The Complete Algorithm}

We now present the complete algorithm by implementing the definitions into the algorithm framework.
Given the partial solution $H$, the candidate edge set of agent $i$ is defined as
\begin{equation*}
    S_i = \{ e\in E\setminus H : f_i(H\cup \{ e \}) > f_i(H) \}.
\end{equation*}
Let function $F$ be the sum of all agents' progress functions, i.e., $F(H) = \sum_{i\in K} f_i(H)$.
Note that $F(H) \in \{ 0,1,\ldots,k(n-1) \}$, and $H$ is a feasible solution if and only if $F(H) = k(n-1)$.
Since the functions $(f_i)_{i\in K}$ are monotone, so is $F$. 
Therefore, $k(n-1) - F(H)$ measures how far an edge set $H$ is away from being feasible.
Selecting the edge that appears in most candidate edge sets can be equivalently described as picking $e\in E\setminus H$ with maximum \emph{marginal contribution} $F( H \cup \{e\} ) - F(H)$ to the current partial solution $H$.
The formal description of the algorithm can be found in \cref{alg:approx}.

\begin{algorithm}[htb] 
\caption{Multi-Round Plural Voting Algorithm}
\label{alg:approx}
\begin{algorithmic}[1]
\REQUIRE A connected undirected graph $G(V,E)$; Agent set $K$ with preferences $\vpreceq=(\preceq_1,\ldots,\preceq_k)$.
\ENSURE A subgraph $H\subseteq E$.
\STATE $H\leftarrow\emptyset$.
\WHILE{$F(H) < k(n-1)$}
\label{line:sub:while:start}
\STATE $e^*\leftarrow\argmax_{e\in E\setminus H} \{ F(H\cup\set{e})-F(H) \}$.
\label{line:sub:element}
\STATE $H\leftarrow H\cup\set{e^*}$.
\ENDWHILE
\label{line:sub:while:end}
\RETURN $H$.
\end{algorithmic}
\end{algorithm}

The analysis to show that \cref{alg:approx} computes an $\cO(\ln k)$-approximation solution to \MMCP/ polynomially consists of the following three steps.
\begin{enumerate}
    \item We show that given any partial solution $H\subseteq E$, the value of $F(H)$ can be computed in polynomial time.
    Since $F(H) = \sum_{i\in K} f_i(H)$, it suffices to provide an oracle for each agent $i$ that answers the value of $f_i(H)$ for each queried set $H\subseteq E$. The following lemma is proved in \cref{sec:alg:computing_f}.
    
    \begin{restatable}{lemma}{computation}
    \label{lem:compute_f}
    Given the preference $\preceq_i$ of agent $i$ and a subset of edges $H\subseteq E$, $f_i(H)$ can be computed in polynomial time.
    \end{restatable}
    
    \item We show that the function $F$ is submodular.
    Again, since $F = \sum_{i\in K} f_i$, it suffices to show the submodularity of each function $f_i$. The following lemma is proved in \cref{sec:alg:submodular_f}.
    
    \begin{restatable}{lemma}{submodular}
    \label{lem:submodular}
        For each agent $i\in K$, the progress function $f_i$ is submodular.
    \end{restatable}
    
    \item We show that \cref{alg:approx} returns a solution $H$ with $|H| \leq \cO(\ln k)\cdot |H^*|$.
    Given the monotonicity and submodularity of $F$, we borrow a result in~\cite{DBLP:journals/combinatorica/Wolsey82} to prove \cref{thm:submodular_algo}.
    
    \begin{lemma}[\cite{DBLP:journals/combinatorica/Wolsey82}]\label{lem:submodular_cover}
    Given any integer-valued monotone submodular function $F: 2^E \to \mathbb{Z}$ with $F(\emptyset) = 0$, the algorithm that repeatedly includes the element with maximum marginal value computes a set $H$ such that (where $d = \max_{e\in E} \{ F(e) \}$)
    \begin{equation*}
        |H| \leq \left(\sum_{i=1}^{d} \frac{1}{i}\right) \cdot \min \{|S|: F(S) = F(E), S\subseteq E\}.
    \end{equation*}
    
    \end{lemma}
\end{enumerate}

Given the above three lemmas, we prove \cref{thm:submodular_algo} as follows.

\begin{proofof}{\cref{thm:submodular_algo}}
    Regarding the correctness of the algorithm, since the returned set $H$ satisfies $F(H) = k(n-1)$, the solution $H$ is feasible.
    By definition of the progress function (\cref{def:progress_function}), $f_i$ is integer-valued, monotone and $f_i(\emptyset) = 0$ for all $i\in K$.
    By \cref{lem:submodular}, $f_i$ is submodular.
    Therefore the function $F = \sum_{i\in K} f_i$ is integer-valued, monotone, submodular and $F(\emptyset) = 0$.
    Moreover, since $f_i(e) \leq 1$ for any agent $i$, we have $\max_{e\in E} \{ F(e) \} \leq k$. 
    Hence by \cref{lem:submodular_cover}, we have
    \begin{equation*}
        |H| \leq \left( \sum_{i=1}^{k} \frac{1}{i} \right) \cdot |H^*| < (\ln k + 1)\cdot |H^*|.
    \end{equation*}
    
    Regarding the running time of the algorithm, it suffices to show that there are a polynomial number of while-loops, since by \cref{lem:compute_f} $F(H)$ can be computed in polynomial time.
    As $F$ is submodular and integer-valued, when $F(H) < k(n-1)$, there must exists $e$ such that $F(H\cup \{e\}) - F(H) \geq 1$.
    Hence in each round, the element $e^*$ selected in line~\ref{line:sub:element} of \cref{alg:approx} increases $F(H)$ by at least one.
    Consequently, there are at most $k(n-1)$ rounds, and the algorithm runs in polynomial time.
\end{proofof}

\subsection{Oracle for the Progress Function}
\label{sec:alg:computing_f}

In this subsection, we prove \cref{lem:compute_f} by giving an algorithm (\cref{alg:compute-f}) for the computation of $f_i(H)$.
We restate \cref{lem:compute_f} for completeness.

\computation*

Recall that agent $i$'s preference $\preceq_i$ generates an ordered partition $\cE(\preceq_i)=(E_i^1,\ldots,E_i^{\kappa_i})$ of the edge set $E$.
Given any agent $i$ and any edge set $H$, \cref{alg:compute-f} aims to find an MST $T$ under $\preceq_i$ such that $T$ uses a maximum number of edges from $H$.
To do that, we use Kruskal's algorithm and break ties in favor of $H$.
In other words, when running Kruskal's algorithm, within each equivalence class $E_i^j$, \cref{alg:compute-f} gives a higher priority to edges in $E_i^j\cap H$.

\begin{algorithm}[htb] 
\caption{Degrade Algorithm}
\label{alg:compute-f}
\begin{algorithmic}[1]
\REQUIRE A connected undirected graph $G(V,E)$; Agent $i$ with preference $\preceq_i$; An edge set $H\subseteq E$.
\ENSURE The value of $f_i(H)$.
\STATE Let $\cE_i \gets (E_i^1,\ldots,E_i^{\kappa_i})$ be the ordered partition generated by the preference $\preceq_i$.
\STATE Let $\cE'_i \gets (E_i^1\cap H, E_i^1\setminus H, \ldots, E_i^{\kappa_i}\cap H, E_i^{\kappa_i}\setminus H)$ and remove empty equivalence classes.\\
\label{line:f:degrade}
\STATE Let $\preceq'$ be the preference such that $\cE(\preceq')=\cE_i'$.
\STATE $T\leftarrow\kru(G,\preceq')$ \label{line:T-compute-f}
\STATE $f_i(H)\leftarrow \abs{T \cap H}$.
\RETURN $f_i(H)$.
\end{algorithmic}
\end{algorithm}

\begin{proofof}{\cref{lem:compute_f}}
We show that \cref{alg:compute-f} (which clearly runs in polynomial time) computes an MST $T\in \mst(G,\preceq_i)$ that has maximum overlap with $H$ in line~\ref{line:T-compute-f}, which by definition implies $f_i(H) = |T\cap H|$, and proves \cref{lem:compute_f}.
Note that $T\in \mst(G,\preceq_i)$ holds because \cref{alg:compute-f} is indeed Kruskal's algorithm (\cref{alg:kruskal}), but with a carefully designed way to break ties. 

Let $|T\cap H| = t$.
For the sake of contradiction, suppose there exists another MST $T' \in \mst(G,\preceq_i)$ such that $|T’\cap H| \geq t+1$.
We construct a weight function $w$ under which both $T$ and $T'$ are MSTs, but with different total weights, which is a contradiction.
Fix any weight function $w_i$ that is consistent with $\preceq_i$.
We define another weight function $w$ as follows, where $\epsilon>0$ is arbitrarily small, e.g., $\epsilon \leq 1/m\cdot |w_i(R)-w_i(R')|$ for any $R,R'\subseteq E$ with $w_i(R)\ne w_i(R')$.
\begin{equation*}
    w(e)=
    \begin{cases}
    w_i(e), & e\notin H, \\
    w_i(e) - \epsilon, &e\in H.
    \end{cases}
\end{equation*}

Observe that $w$ is consistent with $\preceq'_i$, as we give edges in $H$ a (very slightly) higher priority.
Thus by line~\ref{line:T-compute-f} of \cref{alg:compute-f} and \cref{lem:mst_prec_weight}, we have $T\in \mst(G,\preceq'_i)=\mst(G,w)$.
On the other hand, since $T'\in \mst(G,w_i)$ is the MST that has maximum overlap with $H$, we have $T' \in \mst(G,w)$ (because the only difference between $w$ and $w_i$ is that the weight of every edge in $H$ is decreased by $\epsilon$).
Therefore, both $T$ and $T'$ are MSTs under $w$.
Recall that both $T$ and $T'$ are MSTs under $w_i$, which gives $w_i(T) = w_i(T')$.
However, we have $w(T) = w_i(T) - t\cdot \epsilon$ while $w(T') \leq w_i(T') - (t+1)\cdot \epsilon$, which is a contradiction.
\end{proofof}

\subsection{Submodularity of the Progress Function}
\label{sec:alg:submodular_f}

In this subsection, we prove \cref{lem:submodular}, which is restated as follows.

\submodular*

Consider any agent $i\in K$.
Before proving~\cref{lem:submodular}, we first show some properties of the function $f_i$.
For any subset $H\subseteq E$, let $\mst_i^H:=\set{T \in \mst_i \mid f_i(H)=\abs{T\cap H} }$ be the set of minimum spanning trees that contain exactly $f_i(H)$ edges in $H$. We observe the following simple facts.

\begin{observation}
\label{obs:0-or-1}
For all edge $e\in E\setminus H$, we have $f_i(H\cup \{e\}) - f_i(H) \in \{0,1\}$.
\end{observation}

\begin{observation}
\label{obs:increment}
For all edge $e\in E\setminus H$, $f_i(H\cup \{e\}) - f_i(H) = 1$ if and only if edge $e\in T$ for some spanning tree $T\in \mst_i^H$.
\end{observation}

\begin{observation}
\label{obs:strictly_increase}
For all edge $e\in H$, $f_i(H\setminus \{e\}) = f_i(H) - 1$ if and only if $e\in T$ for all $T \in \mst_i^{H}$. 
\end{observation}

\begin{figure}[tb]
    \centering
    \includegraphics[width=12.5cm]{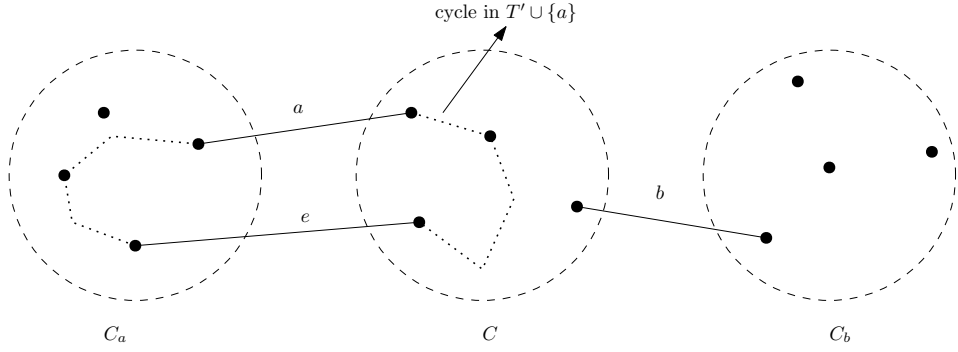}
    \caption{An illustration for the proof of \cref{lem:submodular}. After removing edge $a$ and $b$ from $T$, we have three connected components $C_a$, $C$ and $C_b$. Now, we consider an MST $T'\in \mst_i^S$. After adding edge $a$ to $T'$, there will be a cycle in $T'\cup\set{a}$ and thus there exists an edge $e$ other than $a$ in both the cycle and the cut $\cC(C_a,C\cup C_b)$. By the property of the minimum spanning tree, we prove that $e\sim_i a$.}
    \label{fig:submodular}
\end{figure}

\begin{proofof}{\cref{lem:submodular}}
Consider any agent $i\in K$.
We prove that $f_i$ is submodular by contradiction. 
Suppose that $f_i$ is not submodular.
By definition of submodularity, there must exist a subset $S\subsetneq E$ and two edges $a,b\in E\setminus S$ such that 
\begin{equation}\label{eq:submodular}
    f_i(S\cup \{a\}) -f_i(S) < f_i(S\cup \{a,b\}) - f_i(S\cup \{b\}).
\end{equation}

By \cref{obs:0-or-1}, the above inequality implies $f_i(S\cup \{a\}) = f_i(S)$ and $f_i(S\cup \{a,b\}) = f_i(S\cup \{b\})+1$. 
Reordering \cref{eq:submodular} gives
\begin{equation}\label{eq:submodular2}
    f_i(S\cup \{b\}) -f_i(S) < f_i(S\cup \{a,b\}) - f_i(S\cup \{a\}),
\end{equation}
which implies that $f_i(S\cup \{b\}) = f_i(S)$ and $f_i(S\cup \{a,b\}) = f_i(S\cup \{a\})+1$. 
Thus, we have
\begin{equation}\label{eq:submodular_main}
    f_i(S) = f_i(S\cup \{a\}) = f_i(S\cup \{b\}) = f_i(S\cup \{a,b\}) -1.
\end{equation}

By \cref{obs:increment}, the first equality of \cref{eq:submodular_main} implies that $a\notin T$ for all $T\in \mst_i^S$.
By \cref{obs:strictly_increase}, the last two equations of \cref{eq:submodular_main} implies that for all $T \in \mst_i^{S\cup \{a,b\}}$, we have $\{a,b\} \subseteq T$.
Fix any $T\in \mst_i^{S\cup \{a,b\} }$, and consider the graph $G'=(V,T\setminus \{a,b\})$.
Note that $G'$ has three connected components, say with nodes $C$, $C_a$ and $C_b$, where $a\in \cC(C_a, C\cup C_b)$ and $b\in \cC(C_a\cup C, C_b)$ (see \cref{fig:submodular}).
Here we use $\cC(X,V\setminus X) = E\cap (X\times (V\setminus X))$ to denote the set of cut edges between $X$ and $V\setminus X$.
Observe that 
\begin{itemize}
    \item[(i)] there does not exist $e\in \cC(C_a, C\cup C_b)$ such that $e \prec_i a$ because otherwise $T$ is strictly worse than $T \cup \{ e \} \setminus \{a\}$ and is not an MST under $\preceq_i$;
    \item[(ii)] for all $e\in \cC(C_a, C\cup C_b)\cap (S\cup \{ b \})$, we must have $a \prec_i e$ because otherwise we obtain another MST $T^* = T \cup \{ e \} \setminus \{a\}$ such that $T^* \in \mst_i^{S\cup \{a,b\}}$, which contradicts the fact that $\{a,b\}\subseteq T$ for all $T\in \mst_i^{S\cup \{a,b\}}$.
\end{itemize}

Fix any $T'\in \mst_i^S$ and consider the graph $G'' = (V,T'\cup \{a\})$.
Recall that $a\notin T'$ and thus $G''$ contains a cycle.
Let $e$ be any edge other than $a$ from cut $\cC(C_a, C\cup C_b)$ that also appears in the cycle (see \cref{fig:submodular}).
Note that we cannot have $a\prec_i e$ because otherwise the MST $T'$ can be improved (by removing $e$ and including $a$), which is a contradiction.
By statement (i) from above, we cannot have $e\prec_i a$, and thus $a\sim_i e$;
by statement (ii) from above, we have $e\notin S$.
Therefore, $T^* = T' \cup \{a\} \setminus \{e\}$ is another MST, and contains the same number of edges from $S$ as $T'$, i.e., $T^*\in \mst_i^S$.
However, this is again a contradiction because we have proved that for all $T\in \mst_i^S$, $a\notin T$.
\end{proofof}

\section{Extensions}
\label{sec:extension}

In this section, we present some extensions of the model, and show that our algorithm applies to generalizations of the problem with submodular cost functions and maximum Matroid rank constraints.

\subsection{Submodular Cost Function}

This subsection considers the more general case of \MMCP/ where we have a submodular cost function $c:2^E\to\R_{\geq 0}$ defined on the edge set $E$.
We aim to select a subgraph $H\subseteq E$ with the minimum cost $c(H)$ such that $H$ contains an MST for every agent, i.e., $\mst(G,\preceq_i)\cap\mst(H,\preceq_i)\ne\emptyset$ for all $i\in K$.
We refer to such a problem as the {\em submodular multiagent MST cover} problem (SMMCP). 
We show that \cref{alg:approx} can be extended to SMMCP by choosing the edge with the highest ``price-performance" ratio.
Namely, in each round, we choose the element $e^*$ by the following rule:
\begin{equation*}
    e^* \leftarrow \argmax_{e\in E\setminus H}\left\{\frac{F(H\cup\set{e})-F(H)}{c(e)}\right\}.
\end{equation*}

See \cref{alg:mcsc} for a formal description.
Basically, the algorithm replaces line \ref{line:sub:element} in \cref{alg:approx} with the above rule.

\begin{algorithm}[htb] 
\caption{Algorithm for submodular multiagent MST cover}
\label{alg:mcsc}
\begin{algorithmic}[1]
\REQUIRE A connected undirected graph $G(V,E)$; Agent set $K$ with preferences $\vpreceq =(\preceq_1,\ldots,\preceq_k)$; Submodular cost function $c:2^{E}\to\R_{\geq 0}$.
\ENSURE A subgraph $H\subseteq E$.
\STATE $H\leftarrow\emptyset$.
\WHILE{$F(H)<k(n-1)$}
\STATE $e^* \leftarrow \argmax_{e\in E\setminus H}\left\{ \frac{1}{c(e)}\cdot (F(H\cup\set{e})-F(H)) \right\}$. 
\STATE $H\leftarrow H\cup\set{e^*}$.
\ENDWHILE
\RETURN $H$.
\end{algorithmic}
\end{algorithm}

Let $\cH^*$ be the set of all optimal solutions, i.e., $H$ is a subgraph cover with minimum $c(H)$ for all $H\in\cH^*$.
Let $\gamma=\min_{H\in\cH^*}\left\{\frac{\sum_{e\in H}c(e)}{c(H)}\right\}$ be the minimum curvature of the optimal solution.

\begin{theorem}
For any instance of SMMCP, \cref{alg:mcsc} computes an $\cO(\gamma\cdot\ln k)$-approximation solution in polynomial time, where $k$ is the number of agents.
\label{thm:submodular_cost}
\end{theorem}

Consider a special case that the cost function is additive. Refer to this case as the {\em weighted multiagent MST cover} problem (WMMCP).
In WMMCP, we have $\gamma = 1$ because $\frac{\sum_{e\in S}c(e)}{c(S)}=1$ for any $S\subseteq E$.
Since WMMCP generalizes MMCP, the lower bound $(1-o(1))\ln k$ remains valid.
Thus, \cref{alg:mcsc} is asymptotically optimal for WMMCP; see \cref{thm:additive_cost}.

\begin{corollary}
For any instance of WMMCP, \cref{alg:mcsc} computes an $\cO(\ln k)$-approximation solution in polynomial time, where $k$ is the number of agents.
Moreover, \cref{alg:mcsc} is asymptotically optimal.
\label{thm:additive_cost}
\end{corollary}

Next, we prove \cref{thm:submodular_cost}.
By \cref{lem:compute_f} and \cref{lem:submodular}, we know that the function $F$ is submodular and can be computed in polynomial time.
We borrow a result in \cite[Theorem 2.1]{DBLP:journals/coap/WanDPW10}.

\begin{lemma}[\cite{DBLP:journals/coap/WanDPW10}]
Given any integer-valued monotone submodular function $F:2^E\to\Z$ with $F(\emptyset)=0$ and monotone submodular cost function $c:2^{E}\to\R_{\geq}$, the algorithm that repeatedly includes the element with the maximum ratio of marginal value and cost computes a set $H$ such that 
$$
c(H) \leq \left( \rho \cdot \sum_{i=1}^{d}\frac{1}{i} \right) \cdot \min\set{c(S):F(S)=F(E),S\subseteq E}
$$
where $\rho=\min_{S\in\cO^*}\left\{\frac{\sum_{e\in S }c(e)}{c(S)}\right\}$ ($\cO^*$ is the set of optimal solutions) and $d=\max_{e\in E}\{ F(e)\}$.
\label{lem:submodular_cost_cover}
\end{lemma}

\begin{proofof}{\cref{thm:submodular_cost}}
The correctness and running time of the algorithm directly follow from the proof of \cref{thm:submodular_algo}.
Then by \cref{lem:submodular_cost_cover}, we have
    \begin{equation*}
        c(H) \leq \left( \min_{H\in\cH^*}\left\{\frac{\sum_{e\in H}c(e)}{c(H)}\right\}\cdot \sum_{i=1}^{k} \frac{1}{i} \right) \cdot c(H^*) \leq \left( \gamma \cdot (\ln k + 1) \right)\cdot c(H^*).
    \end{equation*}
Therefore, the solution $H$ is an $\cO(\gamma\cdot \ln k)$ approximation. 
\end{proofof}

\subsection{Matroid Rank Constraints}

In this subsection, we consider a more general model in which there is a common element set $E$, and each agent $i\in K$ has a Matroid $(E,\cI_i)$ defined on $E$.
As in the previous subsection, there is a submodular cost function $c: 2^E \rightarrow \R_{\geq 0}$ defined on the element set.

\begin{definition}[Matroid]
\label{def:Matroid}
    A set system $\cI \subseteq 2^E$ defined on element set $E$ is call a \emph{Matroid} if
    \begin{itemize}
        \item (Hereditary Property) for all $S\in \cI$ and $T\subseteq S$, we have $T\in \cI$;
        \item (Augmentation Property) for all $S\in \cI$, $T\in \cI$ and $|S|>|T|$, there exists an element $e\in S\setminus T$ such that $T\cup \{e\}\in \cI$.
    \end{itemize}
    We call a set $S$ \emph{independent} if $S\in \cI$.
    We denote this Matroid by $(E,\cI)$.
\end{definition}

\begin{definition}[Rank]
    Given a set of element $S\subseteq E$, the rank of $S$ under Matroid $(E,\cI)$ is the size of the largest subset of $S$ that is independent, e.g.,
    \begin{equation*}
        \rank(S) = \max\{ |T|:T\subseteq S, T\in \cI \}.
    \end{equation*}
\end{definition}

We denote by $\rank_i$ the rank function of Matroid $(E,\cI_i)$.
The problem is to compute a subset of elements $S$ with minimum total cost $c(S)$, such that $\rank_i(S) = \rank_i(E)$ for all $i\in K$.
We assume that an oracle that answers a rank query in polynomial time is given for each agent.
We call the above constraints the \emph{maximum Matroid rank} (MMR) constraints, and refer to this more general problem as MCSC with MMR constraints.
We show that
\begin{itemize}
    \item the MST constraint is a special case of the MMR constraint;
    \item our $\cO(\gamma\cdot \ln k)$ approximation algorithm can be extended to the problem with MMR constraints.
\end{itemize}

\subsubsection{The Matroid Defined by MST}

In this part, we show that the MSC problem with MST constraint is a special case of the problem with maximum Matroid rank constraint.
Indeed, as we have shown in \cref{sec:approx} the progress function $f_i(\cdot)$ of each agent $i$ is a submodular function with binary marginals, which implies that there exists a Matroid whose rank function is equivalent to $f_i(\cdot)$~\cite[Chapter 39]{schrijver2003combinatorial1}.

In the following, we explicitly construct a Matroid (for each agent $i$, under preference $\preceq_i$) to establish this equivalence.
We define the following set system $(E,\cI_i)$, where $\cI_i \subseteq 2^E$, for each agent $i\in K$.
For all $S\subseteq E$, we have $S\in \cI_i$, i.e., $S$ is independent, if and only if there exists a superset $T$ of $S$ that is an MST (under a weight function consistent with the preference $\preceq_i$) of agent $i$.

\begin{lemma}
    For each agent $i\in K$, the set system $(E,\cI_i)$ is a Matroid.
\end{lemma}
\begin{proof}
    We show that the set system $\cI_i$ defined above satisfies the two properties in \cref{def:Matroid}. The hereditary property trivially holds by the definition of $\cI_i$.
    Next, we consider the augmentation property.
    
    Consider any two sets $S_1$ and $S_2$ that are both in $\cI_i$, such that $|S_1| > |S_2|$.
    We show that there exists an element $e^*\in S_1$ such that $S_2\cup\{e^*\}$ is also in $\cI$, i.e., there exists an MST $T$ of agent $i$ such that $S_2\cup\{e^*\}\subseteq T$.
    By definition, there exists a superset $T_1$ of $S_1$ that is an MST.
    
    \begin{claim}[Swap Rule]\label{claim:swap-rule}
    For all $e\in S_2$ and any MST $T$ not containing $e$, there exists an edge $e'\in T$ such that $e'\notin S_2$ and $T\cup \{ e \} \setminus \{e'\}$ is also an MST.
    \end{claim}
    \begin{proof}
        Observe that in the subgraph induced by $T\cup \{e\}$, there exists a cycle $C\subseteq T\cup \{e\}$ containing $e$.
        We show that there must exist another edge $e'$ in the cycle such that $w_{e'} = w_e$.
        
        First, note that for all $e'\in C$ we must have either $e' \prec_i e$ or $e'\sim_i e$, as otherwise removing $e'$ from $T\cup \{e\}$ gives a spanning tree strictly better than $T$, which is impossible.
        On the other hand, by the cut property~\cite[page~856]{schrijver2003combinatorial1} of MST, there exists a cut $F$ in which $e$ has the minimum weight.
        Since $F$ is a cut, at least one more edge from the cycle must appear in $F$. Hence there must exists another edge $e'\in C$ such that $e' \sim_i e$, which implies that $T\cup \{ e \} \setminus \{e'\}$ is also an MST. 
        
        It remains to show that there exists at least one such $e'\in C$ that is not in $S_2$.
        Suppose otherwise, i.e., every edge $e'$ with $e' \sim_i e$ in the cycle belongs to $S_2$.
        In other words, for all edge $\hat{e}\in C\setminus S_2$, we have ${\hat{e}} \prec_i e$.
        Since $S_2$ is independent, there exists an MST $T_2$ that is a superset of $S_2$.
        Note that such an MST can be computed using Kruskal's algorithm, as long as we break ties in favor of $S_2$.
        However, in Kruskal's algorithm, all edges $\hat{e}\in C\setminus S_2$ will be scanned before edges in the same equivalence class as $e$ (which is a subset of $C\cap S_2$).
        For each of the scanned edges, either it is included in $T_2$, or it is discarded because its two endpoints are already connected.
        Then we have a contradiction because for the edge $e^* \sim_i e$ (which is in $C\cap S_2$) that is scanned last in the algorithm, its two endpoints must be connected already, and cannot be included in $T_2$.
    \end{proof}
    
    Given the above claim, we can put edges in $S_2$ to $T_1$ one-by-one, each of which swaps out one edge in the MST that is not in $S_2$.
    Therefore the resulting MST contains all edges in $S_2$, and at least one edge $e^*$ from $S_1$ (because at most $|S_2| < |S_1|$ edges are swapped out). Then by definition, $S_2\cup \{e^*\}$ is an independent set, as claimed.
\end{proof}


\subsubsection{Approximation Algorithm for MMR Constraints}

Now we show that our $\cO(\gamma\cdot \ln k)$ approximation algorithm can be extended to SMMCP with MMR constraints. 
The rank function of every agent can be naturally used to measure how much progress the current solution $H$ has made for her, i.e., the larger $\rank_i(H)$ is, the more progress has been made for agent $i$.
Then, we can extend \cref{alg:approx} to SMMCP with MMR constraints by replacing the progress function with the rank function.
See \cref{alg:matroid} for a formal description.

\begin{algorithm}[htb] 
\caption{Algorithm for SMMCP with MMR constraints}
\label{alg:matroid}
\begin{algorithmic}[1]
\REQUIRE Agent set $K$ with $k$ Matroids $(E,\cI_1),\ldots,(E,\cI_k)$; Submodular cost function $c:2^{E}\to\R_{\geq 0}$.
\ENSURE A subset $H\subseteq E$.
\STATE $H\leftarrow\emptyset$.
\WHILE{$\sum_{i=1}^{k}\rank_i(H)<\sum_{i=1}^{k}\rank_i(E)$}
\STATE $e^* \leftarrow \argmax_{e\in E\setminus H}\left\{\frac{1}{c(e)}\cdot (\sum_{i=1}^{k}\rank_i(H\cup\set{e})-\sum_{i=1}^{k}\rank_i(H)) \right\}$. 
\STATE $H\leftarrow H\cup\set{e^*}$.
\ENDWHILE
\RETURN $H$.
\end{algorithmic}
\end{algorithm}

Let $\cH^*$ be the set of all optimal solutions, i.e., $H$ is a feasible element set with minimum $c(H)$ for all $H\in\cH^*$.
Let $\gamma=\min_{H\in\cH^*}\left\{\frac{\sum_{e\in H}c(e)}{c(H)}\right\}$ be the minimum curvature of the optimal solution.

\begin{theorem}
    For any instance of SMMCP with MMR constraints, \cref{alg:matroid} computes an $\cO(\gamma\cdot\ln k)$-approximation solution in polynomial time, where $k$ is the number of agents.
\label{thm:matroid:submodular_cost}
\end{theorem}


Similar to \cref{thm:additive_cost}, if the cost function $c$ is additive, we have the following result.

\begin{corollary}
\cref{alg:matroid} is an $\cO(\ln k)$-approximation algorithm (and thus asymptotically optimal) when the cost function $c$ is additive.
\label{thm:matroid:additive_cost}
\end{corollary}

Now, we give the proof sketch of \cref{thm:matroid:submodular_cost}.
It is well-known that the Matroid rank function is submodular, e.g., see \cite[Lemma 5.1.3]{lau2011iterative} for a formal proof.
Using the following lemma as replacement of~\cref{lem:submodular}, using the same proof as in that of \cref{thm:submodular_cost}, \cref{thm:matroid:submodular_cost} follows immediately.

\begin{lemma}[\cite{lau2011iterative}]
The rank function $\rank:2^{E}\to\set{0,1,\ldots,\abs{E}}$ of a Matroid $(E,\cI)$ is submodular, i.e., for all $A,B\subseteq E$, we have $\rank(A)+\rank(B)\geq\rank(A\cap B)+\rank(A\cup B)$.
\label{lem:rank_function}
\end{lemma}


\section{Conclusion and Open Problems}
\label{sec:conclusion}

In this paper, we propose the multiagent MST cover problem in which we need to compute a minimum subgraph of an input graph that satisfies a group of $k$ heterogeneous agents, where each agent is satisfied if the subgraph contains an MST under its own weight function. 
When the optimal solution of the problem is a spanning tree, the optimal solution can be computed efficiently. However, in the general case, no algorithm can do better than $((1-o(1))\ln k)$-approximation unless $\comP = \NP$.
We then propose an $\cO(\ln k)$-approximation algorithm, achieving the best possible approximation ratio under the assumption that $\comP \neq \NP$.
Further, our algorithm can be naturally generalized to the case when the MST constraints are generalized to maximum Matroid rank constraints, and the cardinality objective function is generalized to submodular.

Our work leaves several interesting problems open.
For example, one possible direction is to relax the satisfaction constraint of the agents.
In our work, we require that every agent is fully satisfied, e.g., can find an MST in the solution.
It would be interesting to investigate whether the approximation ratio can be improved if we allow the solution to contain only approximate MST for each agent.
However, note that the approximation must be additive, as we can change the reduction from the proof of \cref{thm:hardness} to show that the inapproximability remains the same even if the value of MST is $0$ for all agents in $K$.
Additionally, it would be interesting to study the case when agents' constraints are defined by combinatorial structures other than MST and not subsumed within the class of maximum Matroid rank constraints.

\section*{Acknowledgement}

Chenyang Xu is funded by Science and Technology Innovation 2030 – ``The Next Generation of Artificial Intelligence" Major Project No.2018AAA0100900.
Xiaowei Wu is funded by FDCT (File no. 0143/2020/A3, SKL-IOTSC-2021-2023), the SRG of University of Macau (File no. SRG2020-00020-IOTSC) and GDST (2020B1212030003).
Bo Li is funded by HKSAR RGC (No. PolyU 25211321), NSFC (No. 62102333), CCF-HuaweiLK2022006, and PolyU Start-up (No. P0034420).


\printbibliography

\end{document}